\documentclass{article}
\usepackage{amssymb}

\usepackage{graphicx}
\usepackage{graphics}
\usepackage{subfigure}
\usepackage{latexsym, graphicx, epsfig, amssymb, amsmath}
\newcommand{\BibTeX}{{\rm B\kern-.05em{\sc i\kern-.025em b}\kern-.08em
    T\kern-.1667em\lower.7ex\hbox{E}\kern-.125emX}}
\newcommand{\fixfig}{\vspace{-0.2in}}

\begin{document}

\title{Scheduler Vulnerabilities and Attacks in Cloud Computing}

\date{}

\author{
Fangfei Zhou \and
Manish Goel \and
Peter Desnoyers \and
Ravi Sundaram \\
\\
\{youyou,goelm,pjd,koods\}@ccs.neu.edu\\
\\
College of Computer and Information Science \\
Northeastern University, Boston, USA
}

%\author{Fangfei Zhou}
%\ead{youyou@ccs.neu.edu}
%\author{Manish Goel}
%\ead{goelm@ccs.neu.edu}
%\author{Peter Desnoyers}
%\ead{pjd@ccs.neu.edu}
%\author{Ravi Sundaram\corref{cor1}}
%\ead{koods@ccs.neu.edu}


%\address {
%College of Computer and Information Science,
%Northeastern University, Boston, USA
%}

\maketitle

\begin{abstract}
In hardware virtualization a \emph{hypervisor} provides multiple 
Virtual Machines (VMs) on a single physical system, each executing a separate operating system instance. 
The hypervisor schedules execution of these VMs much as the scheduler in an operating system does, balancing factors such as fairness and I/O performance. As in an operating system, the scheduler may be vulnerable to malicious behavior on the part of users seeking to deny service to others or maximize their own resource usage.

Recently, publically available \emph{cloud computing} 
services such as Amazon EC2 have used virtualization to provide customers with virtual machines running on the 
provider's hardware, typically charging by wall clock time rather than resources consumed. Under this business model, manipulation of the scheduler may allow theft of service at the expense of other customers, rather than merely re-allocating resources within the same administrative domain.

We describe a flaw in the Xen scheduler allowing virtual machines to consume almost all CPU time, in preference to other users,
and demonstrate kernel-based and user-space versions of the attack. 
We show results demonstrating the vulnerability in the lab, consuming as much as 98\% of CPU time regardless of fair share, as well as on Amazon EC2, where Xen modifications protect other users but still allow theft of service. In case of EC2, following the responsible disclosure model, we have reported this vulnerability to Amazon; they have since implemented a fix that we have tested and verified (See Appendix B). We provide a novel analysis of the 
necessary conditions for such attacks, and describe scheduler modifications to eliminate the vulnerability. 
We present experimental results demonstrating the effectiveness of these defenses while imposing negligible overhead.
\end{abstract}

\section*{Keyword}
cloud computing, schedulers, security, virtualization, resource management
%\end{keyword}
%\end{frontmatter}

%\begin{document}
\section{Introduction}
Server virtualization \cite{Barham:xen-and-art-of-virtualization} enables multiple instances of an 
operating system and applications (\emph{virtual machines} or VMs) to run on the same physical hardware, as if each were on its own 
machine. 
Recently server virtualization has been used to provide so-called \emph{cloud computing} services, 
in which customers rent virtual machines running on hardware owned and managed by third-party 
providers. Two such cloud computing services are Amazon's Elastic Compute Cloud (EC2) service and  
Microsoft Windows Azure Platform; in addition, similar services are offered by a number of web hosting providers (e.g. Rackspace's Rackspace Cloud 
and ServePath Dedicated Hosting's GoGrid) and referred to as Virtual Private Servers (VPS). In each of these services 
customers are charged by the amount of time their virtual machine is running (in hours or months), rather than by the amount of CPU time used.

The operation of a hypervisor is in many ways similar to that of an operating system; just as an 
operating system manages access by processes to underlying resources, so too a hypervisor 
must manage access by multiple virtual machines to a single physical machine. In either case the choice of scheduling algorithm will involve a trade-off between factors such as fairness, usage caps and scheduling latency.

As in operating systems, a hypervisor scheduler may be vulnerable to behavior by virtual machines which results in inaccurate or unfair scheduling. Such anomalies and their potential for malicious use have been recognized in the past in operating systems---McCanne and Torek~\cite{McCanne93} demonstrate a denial-of-service attack on 4.4BSD, and more recently Tsafrir~\cite{Tsafrir2007} presents a similar attack against Linux 2.6 which was fixed only recently. 
Such attacks typically rely on the use of periodic sampling or a low-precision clock
to measure CPU usage; like a train passenger hiding whenever the conductor checks tickets, 
an attacking process ensures it is never scheduled when a scheduling tick occurs. 

Cloud computing represents a new environment for such attacks, however, for two reasons. First, the economic model of many services renders them vulnerable to theft-of-service attacks, which can be successful with far lower degrees of unfairness than required for strong denial-of-service attacks. In addition, the lack of detailed application knowledge in the hypervisor (e.g. to differentiate I/O wait from voluntary sleep) makes it more difficult to harden a hypervisor scheduler against malicious behavior.

The scheduler used by the Xen hypervisor 
(and with modifications by Amazon EC2) is vulnerable to such timing-based manipulation---rather than receiving its fair share of CPU resources, a VM running on unmodified Xen using our attack can obtain up to 98\% of total CPU cycles,
regardless of the number of other VMs running on the same core. In addition we demonstrate a kernel module allowing unmodified applications to readily obtain 80\% of the CPU.
The Xen scheduler also supports a non-work-conserving (NWC) mode where each VM's CPU usage is ``capped''; in this mode our attack is able to evade its limits and use up to 85\% of total CPU cycles. The modified EC2 scheduler uses this to differentiate levels of service; it protects other VMs from our attack, but we still evade utilization limits (typically 40\%) and consume up to 85\% of CPU cycles. 

We give a novel analysis of the conditions which must be present for such attacks to 
succeed, and present four scheduling modifications which will prevent this attack without sacrificing 
efficiency, fairness, or I/O responsiveness. We have implemented these algorithms, and present experimental results evaluating them on Xen 3.2.1.

The rest of this paper is organized as follows: Section \ref{sec:background} provides a brief introduction 
to VMM architectures, Xen VMM and Amazon EC2 as background. Section \ref{sec:scheduler} describes the 
details of the Xen Credit scheduler. Section \ref{sec:attack} explains our attacking scheme and presents 
experimental results in the lab as well as on Amazon EC2. Next, Section \ref{sec:fixes} details our 
scheduling modifications to prevent this attack, and evaluates their performance and overhead. 
Section \ref{sec:related} discusses related work, and we 
conclude in Section \ref{sec:conclusion}.

\section{Background}
\label{sec:background}
We first provide a brief overview of hardware virtualization
technology in general, and of the Xen hypervisor and Amazon's Elastic
Compute Cloud (EC2) service in particular.

\subsection{Hardware Virtualization}
Hardware virtualization refers to any system which interposes itself
between an operating system and the hardware on which it executes,
providing an emulated or \emph{virtualized} view of physical
resources. Almost all virtualization systems allow multiple operating
system instances to execute simultaneously, each in its own
\emph{virtual machine} (VM). In these systems a Virtual Machine
Monitor (VMM), also known as a \emph{hypervisor}, is responsible for
resource allocation and mediation of hardware access by the various
VMs.

Modern hypervisors may be classified by the methods of
executing guest OS code without hardware access:
(a) binary emulation and translation, (b) paravirtualization, and (c) hardware virtualization support. Binary emulation executes privileged guest code in software, typically with just-in-time translation for
speed \cite{Adams2006:ASPLOS}. Hardware virtualization
support \cite{AMD} in recent x86 CPUs supports a privilege
level beyond supervisor mode, used by the hypervisor to control guest
OS execution. Finally, paravirtualization allows the guest OS to execute directly in user mode, but provides a
set of \emph{hypercalls}, like system calls in a conventional operating system, which the guest uses to perform privileged functions.

\subsection{The Xen Hypervisor}
Xen is an open source VMM for the x86/x64 platform
\cite{Chisnall:xen-book}. It introduced paravirtualization on the x86, using it to support virtualization of
modified guest operating systems without hardware 
support (unavailable at the time) or the overhead of binary
translation. Above the hypervisor
there are one or more virtual machines or \emph{domains} which use
hypervisor services to manipulate the virtual CPU and perform I/O.

\begin{table}[hbt]
\begin{center}
%\vspace{-.5cm}
{\footnotesize
\begin{tabular}{llcc}
Instance Type & Memory & Cores $\times$ speed & \$/Hr\\
\hline
Small &    1.7GB    &    1 $\times$ 1 & 0.085\\
Large &    7.5      &    2 $\times$ 2 & 0.34\\
X-Large &  15       &    4 $\times$ 2 & 0.68\\
Hi-CPU Med. & 1.7   &    \ \ 2 $\times$ 2.5 & 0.17\\
Hi-CPU X-Large & 7  &    \ \ 8 $\times$ 2.5 & 0.68 \\
\end{tabular}}
\caption{Amazon EC2 Instance Types and Pricing. \textmd{(Fall 2010. Speed is given in ``Amazon EC2 Compute Units''.)}}
\label{tab:ec2}
\end{center}
\end{table}

\subsection{Amazon Elastic Compute Cloud (EC2)}
Amazon EC2 is a commercial service which allows customers to run their
own virtual machine instances on Amazon's servers, for a specified
price per hour each VM is running. Details of the different instance
types currently offered, as well as pricing per instance-hour, are
shown in Table \ref{tab:ec2}.

Amazon states that EC2 is powered by ``a highly customized version of
Xen, taking advantage of virtualization'' \cite{Amazon2008:EC2}. The
operating systems supported are Linux, OpenSolaris, and Windows Server
2003; Linux instances (and likely OpenSolaris) use Xen's
paravirtualized mode, and it is suspected that Windows instances do so as well \cite{Boulton2007:WinPara}.

\section{Xen Scheduling}
\label{sec:scheduler}

In Xen (and other hypervisors) a single virtual machine consists of one or 
more virtual CPUs (VCPUs); the goal of the scheduler is to determine which 
VCPU to execute on each physical CPU (PCPU) at any instant. To do this it 
must determine which VCPUs are idle and which are active, and then from 
the active VCPUs choose one for each PCPU.

In a virtual machine, a VCPU is idle when there are no active processes 
running on it and the scheduler on that VCPU is running its \emph{idle task}. 
On early systems the idle task would loop forever; on more modern ones
it executes a HALT instruction, stopping the CPU in a 
lower-power state until an interrupt is received. On a fully-virtualized 
system this HALT traps to the hypervisor and indicates the 
VCPU is now idle; in a paravirtualized system a direct hypervisor call is 
used instead. When an exception (e.g. timer or I/O interrupt)
arrives, that VCPU becomes active until HALT is invoked again.

By default Xen uses the Credit scheduler~\cite{Cherkasova:Comparison},
an implementation of the classic \emph{token bucket} algorithm
in which credits arrive at a constant rate, are conserved up to a maximum, and are expended during service. Each VCPU
receives credits at an administratively determined rate, and a periodic scheduler tick debits credits from the currently running VCPU. If it has no more credits, 
the next VCPU with available credits is scheduled. Every 3 ticks the scheduler 
switches to the next runnable VCPU in round-robin fashion, and distributes new 
credits, capping the credit balance of each VCPU at 300 credits.
Detailed parameters (assuming even weights) are:

\vspace{0.2cm}
{\footnotesize
\begin{tabular}{p{6cm}l}
%\hline
Fast tick period: & 10ms\\
Slower (rescheduling) tick: &30ms\\
Credits debited per fast tick: &100\\
Credit arrivals per fast tick: &100/N\\
Maximum credits: & 300\\
%\hline
\end{tabular} }
\vspace{0.2cm}
%\vspace{0.3cm}

\noindent where N is the number of VCPUs per PCPU.
The fast tick decrements the running VCPU by 100 
credits every 10\,ms, giving each credit a value of 100\,$\mu$s of CPU time; 
the cap of 300 credits corresponds to 30\,ms, or a full scheduling quantum. 
Based on their credit balance, VCPUs are divided into three states: \emph{UNDER}, with 
a positive credit balance, \emph{OVER}, or out of credits, and \emph{BLOCKED} or halted. 

The VCPUs on
a PCPU are kept in an ordered list, with those in UNDER state 
ahead of those in OVER state; the VCPU at the head of the queue is selected 
for execution. In work conserving mode,
when no VCPUs are in the UNDER state, one in the OVER state will be chosen, 
allowing it to receive more than its share of CPU. In
non-work-conserving (NWC) mode, the PCPU will go idle instead.

\begin{figure}[ht]
%\vspace{-0.8cm}
\begin{center}
\includegraphics[width=0.6\linewidth]{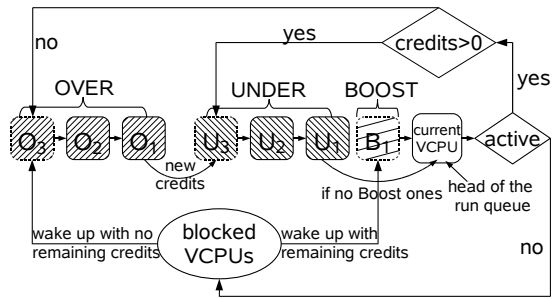}
\end{center}
%\vspace{-0.5cm}
\caption{Per-PCPU Run Queue Structure}
\label{fig:runqueue}
%\vspace{-.2cm}
\end{figure}

The executing VCPU leaves the run queue head in one of two ways: by going idle, or when removed by the scheduler 
while it is still active. VCPUs which go idle enter the BLOCKED state and 
are removed from the queue. Active VCPUs are enqueued after
all other VCPUs of the same state---OVER or UNDER---as shown in Figure~\ref{fig:runqueue}. The basic credit scheduler accurately distributes resources between CPU-intensive 
workloads, ensuring that a VCPU receiving $k$ credits per 30\,ms epoch will receive 
at least $k/10$\,ms of CPU time within a period of 30N\,ms. This fairness comes at the expense of I/O performance, 
however, as events such as packet reception may wait as long as 
30N\,ms for their VCPU to be scheduled.

To achieve better I/O latency, the Xen Credit scheduler attempts to prioritize such I/O. 
When a VCPU sleeps waiting for I/O it will typically have remaining credits; when it wakes with remaining credits it enters the BOOST state  and may immediately preempt running or waiting VCPUs with lower priorities. If it goes idle again with remaining credits, it will wake again in BOOST priority at the next I/O event.

This allows I/O-intensive workloads to achieve very low latency, consuming little CPU and rarely running out of credits, while preserving fair CPU distribution among CPU-bound workloads, which typically utilize all their credits before being preempted. 
However, as we describe in the following section, it also allows a VM to ``steal'' more than its fair share of CPU time. 

\section{Credit Scheduler Attacks}
\label{sec:attack}

Although the Credit scheduler provides fairness and low I/O latency for 
well-behaved virtual machines, poorly-behaved ones can evade 
its fairness guarantees. In this section we describe the features of the 
scheduler which render it vulnerable to attack, formulate an attack 
scheme, and present results showing successful theft of service both 
in the lab and in the field on EC2 instances

\begin{figure}[ht]
\begin{center}
\includegraphics[width=0.6\linewidth]{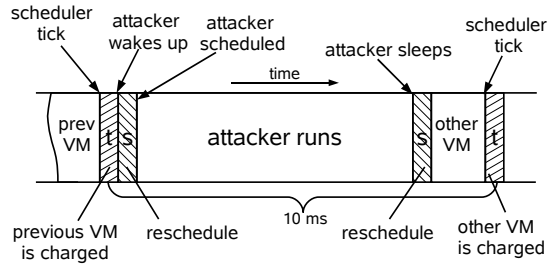}
\end{center}
%\vspace{-1cm}
\caption{Attack Timing}
\label{fig:attack}
\end{figure}
%\vspace{-0.5cm}

\subsection{Attack Description}
\label{sec:scheme}
Our attack relies on periodic sampling as used by the Xen 
scheduler, and is shown as a timeline in Figure
\ref{fig:attack}. Every 10\,ms the scheduler tick fires and schedules
the attacking VM, which runs for 
$10-\varepsilon$\,ms and then calls \emph{Halt()} to briefly go
idle, ensuring that another 
VM will be running at the next scheduler tick. In theory the
efficiency of this attack increases as  
$\varepsilon$ approaches 0; however in practice some amount of timing 
jitter is found, and overly small values of $\varepsilon$ increase the 
risk of the VM being found executing when the scheduling tick arrives. 

When perfectly executed on the non-BOOST credit scheduler, this 
ensures that the attacking VM will never have its credit balance debited. 
If there are $N$ VMs with equal shares, then the N-1 victim 
VMs will receive credits at a total rate of $\frac{N-1}{N}$, and will 
be debited at a total rate of $1$. 

This vulnerability is due not only to the predictability of the
sampling, but to the granularity of the measurement. If the time at
which each VM began and finished service were recorded with a clock
with the same 10\,ms resolution, the attack would still succeed, as the attacker would have a calculated execution time of $0$ on transition to the next VM. 

This attack is more effective against the actual Xen scheduler because of its BOOST priority mechanism. When the attacking VM yields the CPU, 
it goes idle and waits for the next timer interrupt. Due to a lack of information at the VM boundary, however, the hypervisor is unable 
to distinguish between a VM waking after a deliberate sleep 
period---a non-latency-sensitive event---and one waking for e.g. packet reception. The attacker thus wakes in BOOST priority and is able to preempt the currently running VM, so that it can execute for $10-\varepsilon$\,ms out of every 10\,ms scheduler cycle.

\subsection{User-level Implementation}
\label{sec:experiments and analysis}
To examine the performance of our attack scenario in practice, we implement it using both user-level and kernel-based code and evaluate them in the lab and on Amazon EC2. In each case we test with two applications: a simple loop we refer to as ``Cycle Counter'' described below, and the Dhrystone 2.1 \cite{Weicker1988:Dhrystone} CPU benchmark.
Our attack described in Section \ref{sec:scheme} requires
millisecond-level timing in order to sleep before the debit tick and
then wake again at the tick; it performs best either with a tick-less Linux kernel~\cite{Siddha2007:OLS} or with the kernel timer frequency set to 1000\,Hz.

\subsubsection*{Experiments in the lab}

Our first experiments evaluate our attack against unmodified Xen in the lab in work-conserving mode, verifying the ability of the attack to both deny CPU resources to competing ``victim'' VMs, and to effectively use the ``stolen'' CPU time for computation. All experiments were performed on Xen 3.2.1, on a 2-core 2.7\,GHz Intel Core2 CPU.
Virtual machines were 32-bit, paravirtualized, single-VCPU instances with 192\,MB memory, each running Suse 11.0 Core kernel 2.6.25 with a 1000\,Hz kernel timer frequency.

To test our ability to steal CPU resources from other VMs, we implement a ``Cycle Counter'', which performs no useful work, but rather spins using the RDTSC instruction to read the timestamp register and track the time during which the VM is scheduled. The attack is performed by a variant of this code, ``Cycle Stealer'', which tracks execution time and sleeps once it has been scheduled for $10-\varepsilon$ (here $\varepsilon$ = 1\,ms). 

{\footnotesize
\begin{verbatim}
    prev=rdtsc()
    loop:
       if (rdtsc() - prev) > 9ms
           prev = rdtsc()
           usleep(0.5ms)
\end{verbatim}
} 

\begin{figure}[ht]
\centering
\subfigure[CPU (cycle stealer)]
{
  \label{fig:LabTSC}
\includegraphics[width=0.6\linewidth]{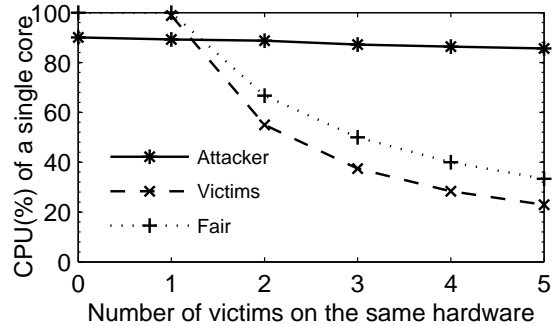}
}
%\hspace{0.5cm}
\subfigure[Application (Dhrystone)]
{
  \label{fig:Labdhry}
  \includegraphics[width=0.6\linewidth]{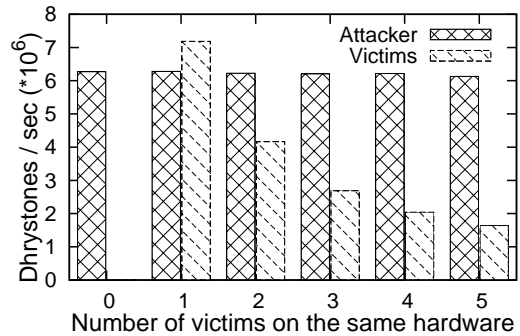}
}
%\vspace{-0.2cm}
\caption{Lab experiments (User Level) - CPU and application performance for attacker and victims.}
\fixfig
\end{figure}

Note that the sleep time is slightly less than $\varepsilon$, as the process will be woken at the next OS tick after timer expiration and we wish to avoid over-sleeping. In Figure~\ref{fig:LabTSC} we see attacker and victim performance on our 2-core test system. As the number of victims increases, attacker performance remains almost constant at roughly 90\% of a single core, while the victims share the remaining core.

\vspace{0.2cm}

To measure the ability of our attack to effectively use stolen CPU cycles, we embed the attack within the Dhrystone benchmark. By comparing the time required for the attacker and an unmodified VM to complete the same number of Dhrystone iterations, we can determine the \emph{net} amount of work stolen by the attacker.

Our baseline measurement was made with one VM running unmodified Dhrystone, with no competing usage of the system; it completed $1.5 \times 10^9$ iterations in 208.8 seconds. 
When running $6$ unmodified instances, three for each core, each  completed the same $1.5 \times 10^9$ iterations in $640.8$ seconds on average---$32.6\%$ the baseline speed, or close to the expected fair share performance of $33.3\%$. With one modified attacker instance competing against $5$ unmodified 
victims, the attacker completed in $245.3$ seconds, running at a speed of $85.3\%$ of baseline, rather than $33.3\%$, with a corresponding decrease in victim performance.
Full results for experiments with 0 to 5 unmodified victims and the modified Dhrystone  attacker are shown in Figure \ref{fig:Labdhry}. 

\begin{figure}[h]
%\vspace{-0.8cm}
\begin{center}
\includegraphics[width=0.6\linewidth]{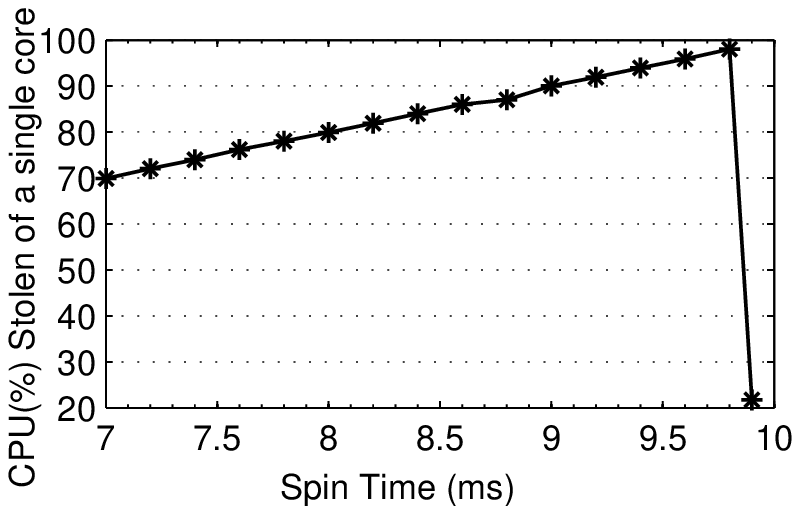}
\end{center}
\fixfig
%\vspace{-0.3cm}
\caption{Lab: User-level attack performance vs. execute times. \textmd{(note - sleep time $\le  10 - $spin time)}}
\label{fig:eposilon}
\end{figure}

In the modified Dhrystone attacker the TSC register is sampled once for each iteration of a particular loop, as described in the appendix; if this sampling occurs too slowly the resulting timing inaccuracy might affect results. To determine whether this might occur, lengths of the compute and sleep phases of the attack were measured. Almost all ($98.8\%$) of the compute intervals were found to lie within the bounds $9 \pm 0.037$\,ms, indicating that the Dhrystone attack was able to attain timing precision comparable to that of Cycle Stealer.

\begin{table}[h]
%\vspace{-0.5cm}
\centering
{ \footnotesize
%\begin{tabular}{cp{0.2cm}cp{0.2cm}cp{0.2cm}c}
\begin{tabular}{c c c c}
&Cycle Stealer & Dhrystones& \% of \\
&(\% of 1 core achieved) & per second & baseline \\ 
\hline
attacker& 81.0 & 5749039 &  80.0\\
victims & 23.4 & 1658553 &  23.1\\
\end{tabular}
} 
\caption{Lab: User-level attack performance in non-work-conserving mode with 33.3\% limit.}
\label{tab:nwc}
\fixfig
\end{table}
\vspace{0.2cm}

As described in Section \ref{sec:scheme}, the attacker runs for a period of length $10-\varepsilon$\,ms and 
then briefly goes to sleep to avoid the sampling tick. A smaller value of $\varepsilon$ increases the 
CPU time stolen by the attacker; however, too small an $\varepsilon$ increases the chance of being charged due to timing jitter. To examine this trade-off
we tested values of $10-\varepsilon$ between $7$ and $9.9$\,ms. 
Figure \ref{fig:eposilon} shows that under lab conditions the peak value was $98\%$ with an execution time of $9.8$\,ms and a requested sleep time of $0.1$\,ms.
When execution time exceeded $9.8$\,ms the attacker was seen by sampling
interrupts with high probability. In this case it received only about $21\%$ of one core, or 
even less than the fair share of $33.3\%$.

%%%%%%%%%%%%%%%%%%%%%%%%

Additional experiments were performed to examine attack performance against the Xen scheduler in non-work-conserving mode. One attacker and 5 victims were run on our 2-core test system, with a CPU cap of 33\% set for each VM; results are presented in Table \ref{tab:nwc} for both Cycle Stealer and Dhrystone attackers, including user-level and kernel-level implementation. 
Here the attacker achieves about $80\%$ of baseline performance with user-level implementation, or slightly less than attacking performance in work-conserving mode. We speculate that in both cases the attacker has a similar chance of being charged by the sampling tick, but that the penalty for being charged is higher in the NWC case.

\subsubsection*{Experiments on Amazon}

We evaluate our attacking using Amazon EC2 Small instances with the following attributes: $32$-bit, $1.7$\,GB memory, $1$ VCPU, running Amazon's Fedora Core $8$ kernel $2.6.18$, with a 1000\,Hz kernel timer. We note that the VCPU provided to the Small instance is described as having ``$1$ EC2 Compute Unit'', 
while the VCPUs for larger and more expensive instances are described as having $2$ or $2.5$ compute units; this indicates that the scheduler is being used in non-work-conserving mode to throttle Small instances. To verify this hypothesis, we ran Cycle Stealer in measurement (i.e. non-attacking) mode on multiple Small instances, verifying that these instances are capped to less than $\frac{1}{2}$ of a single CPU core---in particular, approximately $38\%$ on the measured systems. We believe that the nominal CPU cap for 1-unit instances on the measured hardware is 40\%, corresponding to an unthrottled capacity of 2.5 units.

\begin{figure}[h]
%\vspace{-0.8cm}
\centering
%\subfigure{EC2: Attacker and victim CPU using Cycle Stealer
\subfigure[CPU (cycle stealer)]
{
\includegraphics[width=0.6\linewidth]{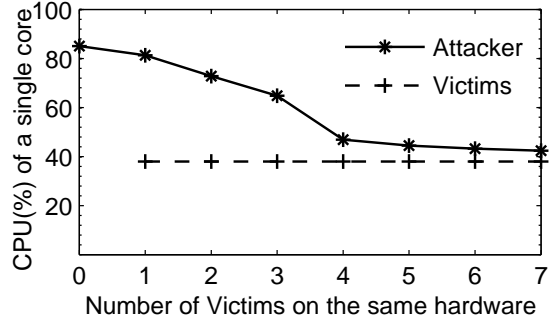}
\label{fig:ATSC}
}
%\hspace{0.5cm}
\subfigure[Application (Dhrystone)]
{
\includegraphics[width=0.6\linewidth]{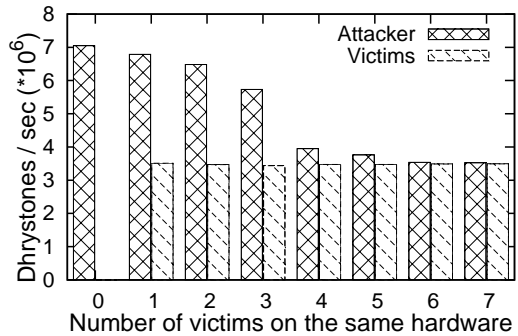}
\label{fig:Adhry}
}
%\vspace{-0.5cm}
\caption{Amazon EC2 experiments - CPU and application performance for
  attacker and victims.}
\end{figure}
%\vspace{-1.2cm}

Additional experiments were performed on a set of 8 Small instances co-located on a single 4-core $2.6$\,GHz physical system provided by our partners at Amazon.\footnote{
This configuration allowed direct measurement of attack impact on co-located ``victim'' VMs, as well as eliminating the possibility of degrading performance of other EC2 customers.}
The Cycle Stealer and Dhrystone attacks measured in the lab were performed in this configuration, and results are shown in Figure \ref{fig:ATSC} and Figure \ref{fig:Adhry}, respectively.
We find that our attack is able to evade the CPU cap of 40\% imposed by EC2 on Small instances, obtaining up to 85\% of one core in the absence of competing VMs.
When co-located CPU-hungry ``victim'' VMs were present, however, EC2 performance diverged from that of unmodified Xen.  As seen in Figures \ref{fig:ATSC} and \ref{fig:Adhry}, co-located VM performance was virtually unaffected by our attack.
Although this attack was able to steal cycles from EC2, it was unable to steal cycles from other EC2 customers.

\subsection{Kernel Implementation}
\label{sec:kernel-level}
Implementing a theft-of-service attack at user level is problematic---it involves modifying and adding overhead to the application consuming the stolen resources, and may not work on applications which lack a regular structure. In contrast, a kernel-module version of this attack allows stolen cycles to be used by arbitrary, unmodified programs.

The attack is implemented as a kernel thread which invokes an OS sleep for $10-\varepsilon$\,ms, allowing user applications to run, and then invokes the \texttt{SCHED\_block} hypercall via the \texttt{safe\_halt} function. In practice $\varepsilon$ must be higher than for the user-mode attack, due to timing granularity and jitter in the kernel.

%\vspace{0.2cm}
{ \footnotesize
\begin{verbatim}
    loop:
       msleep(8);
       safe_halt();
\end{verbatim}
}

\subsubsection{Experimental Results}

\begin{figure}[ht]
%\vspace{-0.8cm}
\centering
\subfigure[CPU (Cycle Counter)]
{
  \label{fig:KTSC}
\includegraphics[width=0.6\linewidth]{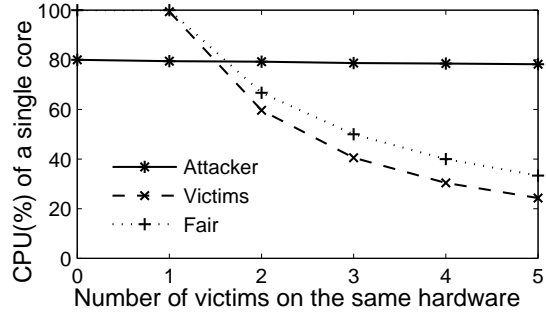}
}
%\hspace{0.5cm}
\subfigure[Application (Dhrystone)]
{
  \label{fig:Kdhry}
  \includegraphics[width=0.6\linewidth]{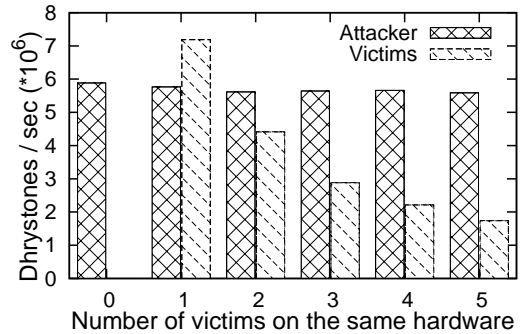}
  }
%\vspace{-0.2cm}
\caption{Lab experiments (Kernel Level) - CPU and application performance for
  attacker and victims.}
\label{fig:K}
\fixfig
\end{figure}
%\vspace{0.2cm}

Lab experiments were performed with one attacking VM, which loads the kernel module, and up to 5 victims running a simple CPU-bound loop. In this case the fair CPU share for each guest instance on our 2-core test system would be $\frac{200}{N}$\% of a single core, where $N$ is the total number of VMs. Due to the granularity of kernel timekeeping, requiring a larger $\varepsilon$, the efficiency of the kernel-mode attack is slightly lower than that of the user-mode attack. In our tests, however, the attacker consumed up to 80.0\% of a single core, with the remaining CPU time shared among the victim instances; results  are shown in Figure \ref{fig:K}. The average amount of CPU stolen by the attacker decreases slightly (from 80.0\% to 78.2\%) as the number of victims increases; we speculate that this may be due to increased timing jitter causing the attacker to occasionally be charged by the sampling tick.

In non-work-conserving mode, with a per-VM CPU cap of 33\%, the attacker was able to obtain 80\% of a single core, as well. In each case the primary limitation on attack efficiency is the granularity of kernel timekeeping; we speculate that by more directly manipulating the hypervisor timer it would be possible to increase efficiency. In addition we note that it often appeared to take seconds or longer for the attacker to synchronize with the hypervisor tick and evade resource caps, while the user-level attack succeeds immediately.

The current implementation of the kernel module does not succeed in stealing cycles on Amazon EC2. Analysis of timing traces indicates a lack of synchronization of the attacker to the hypervisor tick, as seen for NWC mode in the lab, above; in this case, however, synchronization was never achieved. The user-level attack displays strong self-synchronizing behavior, aligning almost immediately to the hypervisor tick; we are investigating approaches to similarly strengthen self-synchronizing in the kernel module.

% LocalWords: paravirtualized unthrottled VCPU VCPUs VMs Xen VM
% LocalWords: Dhrystone timeline hypervisor GHz usleep msleep VM's
% LocalWords: hypercall Suse TSC RDTSC NWC timestamp

\begin{figure}[t]
\begin{center}
\includegraphics[width=0.6\linewidth]{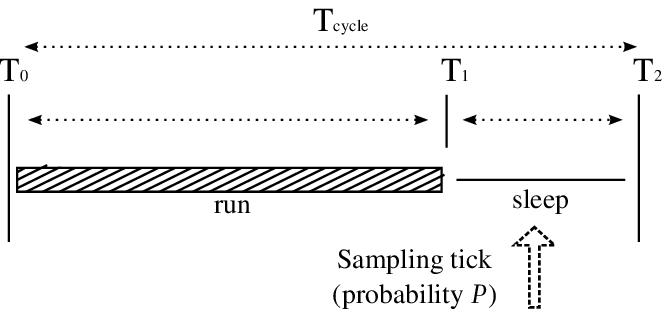}
\end{center}
%\vspace{-0.5cm}
\caption{Attacking trade-offs. \textmd{The benefit of avoiding sampling with probability $P$ must outweigh the cost of forgoing $T_{sleep}$ CPU cycles.}}
\label{fig:stealing}
\fixfig
\end{figure}

\section{Theft-resistant Schedulers}
\label{sec:fixes}

The class of theft-of-service attacks on schedulers which we describe is based on a process or virtual machine voluntarily sleeping when it could have otherwise remained scheduled.
As seen in Figure~\ref{fig:stealing}, this involves a tradeoff---the attack will only succeed if the expected benefit of sleeping for $T_{sleep}$ is greater than the guaranteed cost of yielding the CPU for that time period. If the scheduler is attempting to provide each user with its fair share based on measured usage, then sleeping for a duration $t$  must reduce measured usage by more than $t$ in order to be effective. Conversely, a scheduler which ensures that yielding the CPU will never reduce measured usage more than the sleep period itself will be resistant to such attacks.

This is a broader condition than that of maintaining an unbiased estimate of CPU usage, which is examined by McCanne and Torek~\cite{McCanne93}. Some theft-resistant schedulers, for instance, may over-estimate the CPU usage of attackers and give them less than their fair share. In addition, for schedulers which do not meet our criteria, if we can bound the ratio of sleep time to measurement error, then we can establish bounds on the effectiveness of a timing-based theft-of-service attack.

\subsection{Exact Scheduler}
The most direct solution is the \emph{Exact scheduler}: using a
high-precision clock (in particular, the TSC) to measure actual CPU
usage when a scheduler tick occurs or when a VCPU yields the CPU and
goes idle, thus ensuring that an attacking VM is always charged for exactly the CPU time it has consumed.
In particular, this involves adding logic to the Xen
scheduler to record a high-precision timestamp when a VM begins
executing, and then calculate the duration of execution when it yields
the CPU.
This is similar to the approach taken in e.g. the recent tickless
Linux kernel \cite{Siddha2007:OLS}, where timing is handled by a
variable interval timer set to fire when the next event is due rather
than using a fixed-period timer tick.\footnote{Although Kim et al.
\cite{Kim:Task-aware} use TSC-based timing measurements in their
modifications to the Xen scheduler, they do not address
theft-of-service vulnerabilities.}

\subsection{Randomized Schedulers}
\label{sec:randomized}

An alternative to precise measurement is to sample as before, but on a random schedule. If this schedule is uncorrelated with the timing of an attacker, then over sufficiently long time periods we will be able to estimate the attacker's CPU usage accurately, and thus prevent attack. Assuming a fixed charge per sample, and an attack pattern with period $T_{cycle}$, the probability $P$ of the sampling timer falling during the sleep period must be no greater than the fraction of the cycle $\frac{T_{sleep}}{T_{cycle}}$ which it represents.

\textbf{Poisson Scheduler:} This leads to a Poisson arrival process for sampling, where the expected number of samples during an interval is exactly proportional to its duration, regardless of prior history. This leads to an exponential arrival time distribution,
\[\Delta T=\frac{-ln\emph{U}}{\lambda}\]
where \emph{U} is uniform on (0,1) and \emph{$\lambda$} is the rate parameter of the distribution. We approximate such Poisson arrivals by choosing the inter-arrival time according to a truncated exponential distribution, with a maximum of 30\,ms and a mean of
10\,ms, allowing us to retain the existing credit scheduler structure.
Due to the possibility of multiple sampling points within a 10\,ms period we use a separate interrupt for sampling, rather than re-using or modifying the existing Xen 10\,ms interrupt.

\textbf{Bernoulli Scheduler:}  The discrete-time analog of the Poisson process, the \emph{Bernoulli} process, may be used as an approximation of Poisson sampling. Here we divide time into discrete intervals, sampling at any interval with probability \emph{p} and skipping it with probability \emph{q = 1-p}. 
We have implemented a Bernoulli scheduler with a time interval of 1\,ms, sampling with $p = \frac{1}{10}$, or one sample per 10\,ms, for consistency with the unmodified Xen Credit scheduler.
Rather than generate a timer interrupt with its associated overhead every 1\,ms, 
we use the same implementation strategy as for the Poisson scheduler, generating an inter-arrival time variate and then setting an interrupt to fire after that interval expires.

By quantizing time at a 1\,ms granularity, our Bernoulli scheduler leaves a small vulnerability, as an attacker may avoid being charged during any 1\,ms interval by sleeping before the end of the interval. Assuming that (as in Xen) it will not resume until the beginning of the next 10\,ms period, this limits an attacker to gaining no more than 1\,ms every 10\,ms above its fair share, a relatively insignificant theft of service.

\textbf{Uniform Scheduler:} The final randomized scheduler we propose
is the \emph{Uniform} scheduler, which distributes its sampling uniformly across 10\,ms scheduling intervals.
Rather than generating additional interrupts, or modifying the time at which the existing scheduler interrupt fires, we perform sampling within the virtual machine switch code as we did for the exact scheduler.
In particular, at the beginning of each 10\,ms interval (time $t_0$) we generate
a random offset $\Delta$ uniformly distributed between 0 and 10\,ms.  At each VCPU switch, as well as at the 10\,ms tick, we check to see whether the current time has exceeded $t_0+\Delta$. If so, then we debit the currently running VCPU, as it
was executing when the ``virtual interrupt'' fired at $t_0+\Delta$.

Although in this case the sampling distribution is not memoryless, it is still sufficient to thwart our attacker. We assume that sampling is undetectable by the attacker, as it causes only a brief interruption indistinguishable from other asynchronous events such as network interrupts. In this case, as with Poisson arrivals the expected number of samples within any interval in a 10\,ms period is exactly proportional to the duration of the interval. 

Our implementation of the uniform scheduler quantizes $\Delta$ with 1\,ms granularity, leaving a small vulnerability as described in the case of the Bernoulli scheduler. As in that case, however, the vulnerability is small enough that it may be ignored.
We note also that this scheduler is not theft-proof if the attacker is able to observe the sampling process. If we reach the 5\,ms point without being sampled, for instance, the probability of being charged 10\,ms in the remaining 5\,ms is 1, while avoiding that charge would only cost 5\,ms. 

\subsection{Evaluation}

We have implemented each of the four modified schedulers on Xen 3.2.1. Since the basic credit and priority boosting mechanisms have not been modified from the original scheduler, our
modified schedulers should retain the same fairness and I/O performance properties of the original in the face of well-behaved applications. To verify performance in the face of ill-behaved applications we tested attack performance against the new schedulers; in addition measuring overhead and I/O performance.

\begin{table}[ht]
%\vspace{-0.8cm}
\centering
{\footnotesize
\begin{tabular}{rc}
& CPU(\%) obtained by the attacker \\
Scheduler & (user-level)  \quad (kernel-level) \\\hline
%Unmodified Xen Credit & 85.6  \quad  \quad\quad  \quad abc\\
Xen Credit & 85.6  \quad  \quad\quad  \quad 78.8\\
Exact & 32.9  \quad  \quad\quad  \quad 33.1\\
Uniform & 33.1 \quad  \quad \quad  \quad 33.3 \\
Poisson & 33.0  \quad  \quad\quad  \quad 33.2 \\
Bernoulli & 33.1  \quad  \quad\quad  \quad 33.2\\
\end{tabular} }
\caption{Performance of the schedulers against cycle stealer}
\label{tab:fixes}
%\vspace{-0.3cm}
\end{table}

\begin{table}[ht]
%\vspace{-1.2cm}
\centering
{\footnotesize
\begin{tabular}{rcc}
%& Dhrystones &  \\
%Scheduler & per second & \% of baseline \\ \hline
& Dhrystones/sec (M)& \% of baseline \\
Scheduler & (user)  (kernel) & (user)  (kernel) \\ \hline
%Unmodified Xen Credit & 6128809 & 85.3\\
%Exact & 2317493 & 32.2\\
%Uniform & 2373127 & 33.0\\
%Poisson & 2326233 & 32.4 \\
%Bernoulli & 2336087 & 32.5 \\
%Unmodified Xen Credit & 6.13\quad abc & 85.3\quad abc\\
Xen Credit & 6.13\quad 5.59 & 85.3\quad 77.8\\
Exact & 2.32\quad 2.37 & 32.2\quad 33.0\\
Uniform & 2.37\quad 2.40 & 33.0\quad 33.4\\
Poisson & 2.32\quad 2.38 & 32.4\quad 33.1\\
Bernoulli & 2.33\quad 2.39 & 32.5\quad 33.3 \\
\end{tabular} }
\caption{Performance of the schedulers against Dhrystone}
\label{tab:moddhry}
%\vspace{-.5cm}
\end{table}

\subsubsection{Performance against attack}

In Table \ref{tab:fixes} we see the performance of our Cycle Stealer on the Xen Credit scheduler and the modified schedulers. All four of the schedulers were successful in thwarting the attack: when
co-located with 5 victim VMs on 2 cores on the unmodified scheduler, the attacker was able to
consume 85.6\% of a single-CPU with user-level attacking and 80\% with kernel-level attacking , but no more than its fair share on each of the modified ones. 
(Note that the 85.6\% consumption with user-level attacking in the unmodified case was
limited by the choice of $\varepsilon=1$\,ms, and can increase
with suitably reduced values of $\varepsilon$ as shown in Figure \ref{fig:eposilon}.)

%\begin{figure}[t]
% \begin{center}
%   \includegraphics[width=\columnwidth]{../Figs/Figure7}
% \end{center}
%\caption{Output of modified Dhrystone on different schedulers}
%\label{fig:moddhry}
%\end{figure}

In Table \ref{tab:moddhry} we see similar results for the modified
Dhrystone attacker. Compared to the baseline, the unmodified scheduler allows the 
attacker to steal about 85.3\% CPU cycles with user-level attacking and 77.8\% with kernel-level attacking; while each of the improved schedulers
limits the attacker to approximately its fair share.

\subsubsection{Overhead Measurement}

To quantify the impact of our scheduler modifications on normal execution (i.e. in the absence of attacks) we performed a series of measurements to determine whether application or I/O performance had been degraded by our changes. Since the primary modifications made were to interrupt-driven accounting logic in the Xen scheduler, we examined overhead by measuring performance of a CPU-bound application (unmodified Dhrystone) on Xen while using the different scheduler. To reduce variance between measurements (e.g. due to differing cache line alignment~\cite{Mytkowicz2009:wrongdata}) all schedulers were compiled into the same binary image, and the desired scheduler selected via a global configuration variable set at boot or compile time.

Our modifications added overhead in the form of additional interrupts and/or accounting code to the scheduler, but also eliminated other accounting code which had performed equivalent functions. To isolate the effect of new code from that of the removal of existing code, we also measured versions of the Poisson and Bernoulli schedulers (Poisson-2 and Bernoulli-2 below) which performed all accounting calculations of both schedulers, discarding the output of the original scheduler calculations.

\begin{table}[ht]
%\vspace{-0.8cm}
\centering
{\footnotesize

\begin{tabular}{rp{0.2cm}cp{0.2cm}c}

Scheduler &  & CPU overhead (\%) & & 95\% CI\\ \hline
Exact & & 0.50 & & 0.24 -- 0.76\\
Uniform & & 0.44 & & 0.27 -- 0.61\\
Poisson & & 0.04 & & -0.17 -- 0.24 \\
Bernoulli & & -0.10  & & -0.34 -- 0.15\\
Poisson-2 & & 0.79 & & 0.60 -- 0.98 \\
Bernoulli-2 & & 0.79 & & 0.58 -- 1.00 \\
\end{tabular}} 
\caption{Scheduler CPU overhead, 100 data points per scheduler.}
\label{tab:cpuoverhead}
%\vspace{-1.3cm}
\end{table}

Results from 100 application runs for each scheduler are shown in Table \ref{tab:cpuoverhead}. Overhead of our modified schedulers is seen to be low---well under 1\%---and in the case of the Bernoulli and Poisson schedulers is negligible. Performance of the Poisson and Bernoulli schedulers was unexpected, as each incurs an additional 100 interrupts per second; the overhead of these interrupts appears to be comparable to or less than the accounting code which we were able to  remove in each case. We note that these experiments were performed with Xen running in paravirtualized mode; the relative cost of accounting code and interrupts may be different when using hardware virtualization.

We analyzed the new schedulers' I/O performances by testing the I/O latency between two VMs in two configurations. In configuration 1, two VMs executed  on the same core with no other VMs active, while in configuration 2 a CPU-bound VM was added on the other core. From the first test, we expected to see the performance of well-behaved I/O intensive applications on different schedulers; from the second one, we expected to see that the new schedulers retain the priority boosting mechanism. 

\begin{table}[ht]
%\vspace{-.8cm}
\centering
{\footnotesize
\begin{tabular}{rc}

 & Round-trip delay ($\mu$s) \\
Scheduler \quad & (config. 1) \quad \quad (config. 2) \\ \hline
Unmodified Xen Credit \quad & 53 $\pm$ 0.66 \quad \quad 96 $\pm$ 1.92 \\ 
Exact \quad & 55 $\pm$ 0.61 \quad \quad 97 $\pm$ 1.53 \\
Uniform \quad & 54 $\pm$ 0.66 \quad \quad 96 $\pm$ 1.40 \\
Poisson \quad & 53 $\pm$ 0.66 \quad \quad 96 $\pm$ 1.40 \\
Bernoulli \quad & 54 $\pm$ 0.75 \quad \quad 97 $\pm$ 1.49 \\
\end{tabular} }
\caption{I/O latency by scheduler, with 95\% confidence intervals.}
\label{tab:io}
%\vspace{-1.1cm}
\end{table}

In Table \ref{tab:io} we see the results of these measurements. Differences in performance were minor, and as may be seen by the overlapping confidence intervals, were not statistically significant.

\subsection{Additional Discussion}

\begin{table}[ht]
%\vspace{-0.5cm}
\centering
{\footnotesize
\begin{tabular}{r|llllll} \hline

& Short-run & Long-run & Low & Ease of &Determi&Theft\\
Schedulers&fairness&fairness&overhead&implementation&-nistic&-proof\\ \hline
Exact & $\checkmark$ & $\checkmark$  &  & $\checkmark$ & $\checkmark$ & $\checkmark$ \\ \hline
Uniform &            & $\checkmark$   &  & $\checkmark$ & & \\ \hline
Poisson &  &$\checkmark$ & $\checkmark$ & & & $\checkmark$\\ \hline
Bernoulli & & $\checkmark$ & $\checkmark$& & & \\ \hline
\end{tabular} }
\caption{Comparison of the new schedulers}
\label{tab:compare}
%\vspace{-1cm}
\end{table}

A comprehensive comparison of our proposed schedulers is shown in Table \ref{tab:compare}. The \emph{Poisson} scheduler seems to be the best option in practice, as it has no performance overhead nor vulnerability. Even though it has short-period variance, it guarantees exactly fair share in the long run. The \emph{Bernoulli} scheduler would be an alternative if the vulnerability of up to 1ms is not a concern. The \emph{Uniform} scheduler has similar performance to the Bernoulli one, and the implementation of sampling is simpler, but it has more overhead than Poisson and Bernoulli. Lastly, the \emph{Exact} scheduler is the most straight-forward 
strategy to prevent cycle stealing, with a relatively trivial implementation but somewhat higher overhead.

% LocalWords: quantizing jittering VCPU VCPUs Dhrystone  VMs
% LocalWords: Xen memoryless TSC pdf tickless VM et al timestamp
% LocalWords: paravirtualized virtualization

\section{Related Work}
\label{sec:related}

Tsafrir \emph{et al.} \cite{Tsafrir2007} demonstrate a timing attack on the Linux 2.6 scheduler, allowing an attacking process to appear to consume no CPU and receive higher priority.
McCanne and Torek \cite{McCanne93} present the same cheat attack on 4.4BSD, and develop a uniform randomized sampling clock to estimate CPU utilization. They describe sufficient conditions for this estimate to be accurate, but unlike section \ref{sec:randomized} they do not examine conditions for a theft-of-service attack.

Cherkasova and Gupta \emph{et al.} \cite{Cherkasova:Comparison,Cherkasova:When-virtual-is-harder-than-real} 
have done an extensive 
performance analysis of scheduling in the Xen VMM. They studied I/O performance for the three schedulers: BVT, SEDF and Credit scheduler.
Their work showed that both the CPU scheduling algorithm and the scheduler parameters drastically impact the I/O performance. 
Furthermore, they stressed that the I/O model on Xen remains an issue in resource allocation and accounting among VMs.
Since Domain-0 is indirectly involved in servicing I/O for guest domains, I/O intensive domains may receive excess CPU resources by focusing on the processing resources used by Domain-0 on behalf of I/O bound domains. To tackle this problem, Gupta \emph{et al.} \cite{Gupta:enforcing} introduced the SEDF-DC scheduler, derived from Xen's SEDF scheduler, that charges guest domains for the time spent in Domain-0 on their behalf.

Govindan \emph{et al.} \cite{Govindan:Xen-and-co} proposed a CPU scheduling algorithm as an extension to Xen's SEDF scheduler 
that preferentially schedules I/O intensive domains. The key idea behind their algorithm is to count the number of packages flowing 
into or out of each domain and to schedule the one with highest count that has not yet consumed its entire slice. 

However, Ongaro \emph{et al.} \cite{Ongaro:Scheduling-I/O} pointed out that this scheme is problematic when bandwidth-intensive and 
latency-sensitive domains run concurrently on the same host - the bandwidth-intensive domains are likely to take 
priority over any latency-sensitive domains with little I/O traffic.  They explored the impact of VMM scheduler on I/O performance 
using multiple guest domains concurrently running different types of applications and evaluated 11 different scheduler configurations 
within Xen VMM with both the SEDF and Credit schedulers. They also proposed multiple methods to improve I/O performance.

Weng \emph{et al.} \cite{Weng:hybrid} found from their analysis that Xen's asynchronous CPU scheduling strategy wastes 
considerable physical CPU time. To fix this problem, they presented a hybrid scheduling framework that groups VMs into high-throughput 
type and concurrent type and determines processing resource allocation among VMs based on type. In a similar vein
Kim \emph{et al.} \cite{Kim:Task-aware} presented a task-aware VM scheduling mechanism to improve the performance of I/O-bound tasks 
within domains. Their approach employs gray-box techniques to peer into VMs and identify I/O-bound tasks in mixed workloads.

There are a number of other works on improving other aspects of virtualized I/O performance 
\cite{Cherkasova:Measuring,Liu:VMM-bypass,Willmann:Concurrent,Menon:Diagnosing,Raj:High-performance} and VMM 
security \cite{Rueda,Murray,Jansen}. To summarize, all of these papers tackle problems of long-term fairness between
different classes of VMs such as CPU-bound, I/O bound, etc.

% LocalWords: virtualized VM Ristenpart VMs et al VMM Xen's Weng SEDF 
% LocalWords: Ongaro  Tsafrir Cherkasova Torek McCanne Xen XP Solaris
% LocalWords: pdf pdfs  scheduler's BVT Govindan
\section{Conclusions}
\label{sec:conclusion}

Scheduling has a significant impact on the fair sharing of processing resources among virtual machines
and on enforcing any applicable usage caps per virtual machine. 
This is specially important in commercial services like computing cloud services, where customers who 
pay for the same grade of service expect to receive the same access to resources and providers
offer pricing models based on the enforcement of usage caps. However, the Xen 
hypervisor (and perhaps others) uses a scheduling mechanism which may fail to detect and account 
for CPU usage by poorly-behaved virtual machines, allowing malicious customers to obtain enhanced 
service at the expense of others. The use of periodic sampling to measure CPU usage creates a loophole
exploitable by an adroit attacker.

We have demonstrated this vulnerability in Xen 3.2.1 in the lab, and in Amazon's Elastic Compute 
Cloud (EC2) in the field. Under laboratory conditions, we found that the applications exploiting 
this vulnerability are able to utilize up to 98\% of the CPU core on which they are scheduled, 
regardless of competition from other virtual machines with equal priority and share. Amazon EC2 
uses a patched version of Xen, which prevents the capped amount of CPU resources of other VMs 
from being stolen. However, our attack scheme can steal idle CPU cycles to increase its share, 
and obtain up to 85\% of CPU resources (as mentioned earlier, we have been in discussions with
Amazon about the vulnerability reported in this paper and our recommendations for fixes; they 
have since implemented a fix that we have tested and verified). Finally, we describe 
four approaches to eliminating this 
cycle stealing vulnerability, and demonstrate their effectiveness at stopping our attack in 
a laboratory setting. In addition, we verify that the implemented schedulers offer minor or 
no overhead. 

%\section*{Acknowledgements}
%We would like to thank the team at Amazon Web Services, and in particular Jim Scharf, for 
%their valuable assistance and resources. 
%
%This research was supported in part by a grant from 
%Microsoft Corporation. We would like to make special mention of John Manferdelli of Microsoft
%for invaluable advice and support during the course of this research.

% LocalWords: Xen EC2 VMs hypervisor
% LocalWords: 
% LocalWords: 

% bibliography
%\bibliographystyle{plain-csmin.bst}
\newpage
\bibliographystyle{plain}
\bibliography{refer}

\newpage
\appendix

\section{Modified Dhrystone} 
The main loop in Dhrystone 2.1 is modified by first adding the following constants and global variables. (Note that in practice the CPU speed would be measured at runtime.) 
%\vspace{-0.1in}

{\scriptsize 
\begin{verbatim}
/* Machine-specific values (for 2.6GHz CPU) */
#define ONE_MS   2600000        /* ticks per 10ms */
#define INTERVAL (ONE_MS * 9)   /*           9ms */
#define SLEEP    500            /* 0.5ms (in uS) */

/* TSC counter timestamps */
u_int64_t       past, now;
int             tmp;
\end{verbatim} } 
%\vspace{-0.1in}

The loop in \texttt{main()} is then modified as follows: \vspace{-0.1in}

{ \scriptsize 
\begin{verbatim}
+  int tmp = 0;
   for (Run_Index = 1; Run_Index <= Number_Of_Runs; ++Run_Index)
   {
+      /* check TSC counter every 10000 loops */
+      if( tmp++ >= 10000) {
+          now = rdtsc();
+          if ( now - past >= INTERVAL ) {
+              /* sleep to bypass sampling tick */
+              usleep(SLEEP);
+              past = rdtsc();
+              tmp = 0;
+          }
+      }

       Proc_5();
       Proc_4();
       ...
\end{verbatim} }

\section{Obligations - Legal and Ethical}
Since this research essentially involves a theft of service we include
a brief discussion (in separate paragraphs) of our legal obligations 
under statute and ethical or moral concerns. 

Interaction with computer systems in the United States is covered by
the Computer Fraud and Abuse Act (CFAA) which broadly mandates that
computer system access must be authorized. As is common with any statute
there is considerable ambiguity in the term ``authorization'' and the
complexity derives in large part from case precedents and subsequent
interpretations \cite{Kerr:Cybercrime}. We believe that we were
in full compliance with this statute during the entire course of this
research. We were authorized and paying customers of EC2 and we
did not access any computer systems other than the one we were running
on (for which we were authorized, naturally). All that we did in our
virtual machine was to carefully time our sleeps which, we believe, is completely 
legal.

Once we realized that it was possible to modify a VM to steal cycles we
immediately contacted a few senior executives at Amazon. They, in turn,
put us in touch with members of the team in charge of security for EC2.
We gave them a detailed explanation of the hole we had discovered, along
with the code for our exploit as well as an early draft of this paper.
As is well known \cite{Ristenpart:Cloud} EC2 does not allow its customers
to specify the physical hosts on which their instances are located. We
were concerned that in the course of our experiments we could be stealing 
cycles not only from EC2 but from other unsuspecting customers of EC2 
as well. We requested the security team to give us access to an isolated 
collection of physical hosts on which to conduct our research. Once they
verified our exploit they gave us our own isolated set-up where we were able
to see that our exploit stole unused cycles from EC2 but not from other
VMs co-located on the same host. This means that throughout the entire
course of this research we did not impact other customers. Our exploit
could, at the worst, steal cycles from EC2. The security team then put
up a patched version of EC2 on another separate and isolated set of hosts
and we ran a series of tests confirming that this new version was secure
against our exploit. Up to the point of submission of this article they
had not rolled out this new secure version to the EC2 network. We would
like to point out that rather than go public (e.g. to Slashdot or the
popular press) with our findings we first approached Amazon giving them
the opportunity to fix their exploit. We believe that this amply 
demonstrates our commitment to fully responsible disclosure and 
ethical research that enables a more 
secure Internet both for users as well as infrastructure providers. 
In response to our findings Amazon rolled out a patch and explicitly acknowledged our contribution with the following public testimonial ``Amazon Web Services appreciates research efforts that adhere to the guidelines for responsible disclosure. The Northeastern team has demonstrated its commitment to Internet security by working closely with Amazon Web Services prior to publication of its findings.''

% sigproc.bib is the name of the Bibliography in this case
% ACM needs 'a single self-contained file'!

\end {document}